# Back-reflecting interferometeric sensor based on grating coupler on a planar waveguide

**Anat Demeter-Finzi and Shlomo Ruschin***

*Department of Physical Electronics, School of Electrical Engineering Faculty of Engineering,*

*Tel-Aviv University, Tel-Aviv 69978 Israel*

*[*]Corresponding author: ruschin@eng.tau.ac.il*

*We present a one-port sensor based on a single diffraction grating delineated over a planar optical waveguide. Distinctly to previously reported devices, the grating here is used not only as I/O coupler, but also provides a built-in reference beam which is basically unaffected by the sensing process as manifested in changes of the effective refractive index of the waveguide. The sensing process causes two effects simultaneously: a change in the angle of the out-coupled beam and a change in the phase accumulated by that beam. Both changes can be determined by their conjunction with the reference beam back-diffracted directly by the grating. These two effects are expected to have despair sensitivities, the angle changing effect being coarse and the interferometric phase-change effect being highly sensitive. Sensing simultaneously at two different scales will translate into a large sensing dynamic range.*

# 1. Introduction

The implementation of grating couplers in waveguide sensing structures has attracted considerable attention for many years and proved the ability to provide highly sensitive and rapid detection of analytes. In most reported cases, the sensing mechanism was based on the dependence of the grating couplers' functionality on the specific properties of the underlying waveguide (e.g. the effective refractive index). The sensing mechanism translated into changes of the transmission and/or reflectivity of the sensor [1-4]. Another family of sensors based on waveguide channels is centered on interference effects: an interferometric structure (e.g. Mach-Zehnder type or Young) [5-9], is delineated into an Integrated Optic circuit containing a reference path and a sensing path. A single coherent light wave is there divided into these two paths and suitably recombined. In the sensing arm, the changes in effective refractive index give raise to changes in the accumulated phase, which translate into output power following coherent recombination. Interferometric waveguide sensors are considered to be among the most sensitive, having demonstrated resolutions of the order of $10^{-4}$ -$10^{-6}$ RIU with potential of attaining even higher resolving powers when dispersion effects are suitable harnessed [10,11]. Interferometric sensors however suffer of drawbacks, namely low dynamic range and phase ambiguity. The sensitivity range of these sensors can be tailored to a predetermined range, but in general, they are not suitable for detecting changes at dissimilar scales, unless further complications in design or measurement techniques are involved [12]. The device described in the present paper, solves in a simple way the issue of enhancing the dynamic range and still retains the high sensitivity of interferometric sensors. It does so by combing the two effects mentioned above, namely two-wave interference and the dependence of grating coupling conditions on the physical properties of the waveguide. Furthermore, being based on a single I/O grating, it greatly simplifies the design, fabrication and packaging of the device.

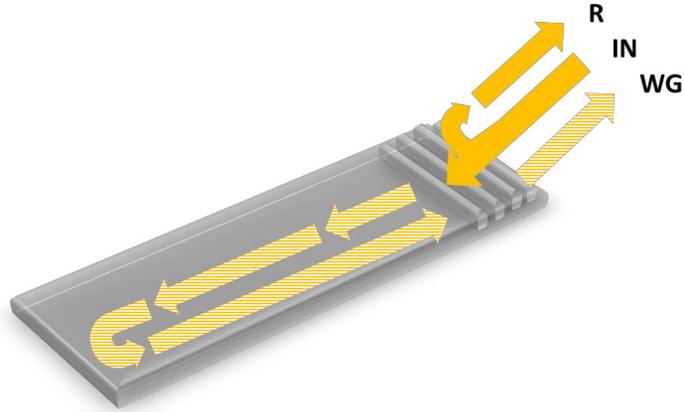

Fig. 1. Schematic drawing of the proposed sensor. The sensor has a single I/O coupler. The out-coupled (*WG*) and the directly diffracted (*R*) beams have the same angle and interference between them takes place at the far-field.

Distinctly from previous work, the sensor presented here measures simultaneously changes in both the coupling angle and the phase accumulated along the waveguide by the interference between an out-coupled beam and a reference beam out-diffracted directly by the grating. The scheme of our sensor is outlined in Fig. 1: A beam with intensity $I_{IN}$ is directed from the air towards a grating at a nearly Littrow configuration. Part of the incident power reflects back and part is coupled into an underlying waveguide, propagates and is back-reflected at the waveguide's end by simple Fresnel reflection or by a dedicated Bragg reflector. Upon its return, the reflected light (*WG*) is coupled-out by the same I/O grating and recombines with the directly back-diffracted beam (*R*). As a result, both outgoing beams overlap and interference is expected between them. Under different concentrations of the sensed material and therefore different values of the waveguide's effective refractive index, we will observe changes in the overlap range of the beams and displacement of the interference fringes. We demonstrate these effects by simulations based on a porous silicon waveguide.

The rest of the paper is organized as follows: In section 2, we summarize the necessary theoretical relationships governing grating couplers and diffraction gratings. We explain about the interferometer structure, the conditions that need to apply in order to attain overlapping between the interfering beams, and we analyse the angular spectrum at the far field. In section 3, we present simulation results and quantitative expected results for a specific structure based on porous silicon, showing the changes in the far-field pattern caused by the changes in the waveguide's effective refractive index. Finite difference time domain simulations are also presented to verify our theoretical model. Conclusions are summarized in section 4. In the Appendix, we deal with the effect of eventual inaccuracies in the fabrication process and measuring procedures on the nominal conditions required for the proper function of the interferometer, and we suggest a solution in order to minimize these effects based on fabrication parameter tolerance considerations.

## 2. Theory

### 2.1 Basic theory and operational principle

Fig. 2(a) displays the basic configuration under discussion, in which the two interfering beams are back-directed into the source direction (Littrow type). Fig. 2(b) represents a variant of that configuration (non-Littrow), where the two out coming beams still overlap but are not back-reflected into the source direction. The choice of either one of these schemes would follow rather practical consideration. Our present analysis is general and encompasses both cases that basically differ in the choice of diffraction orders.

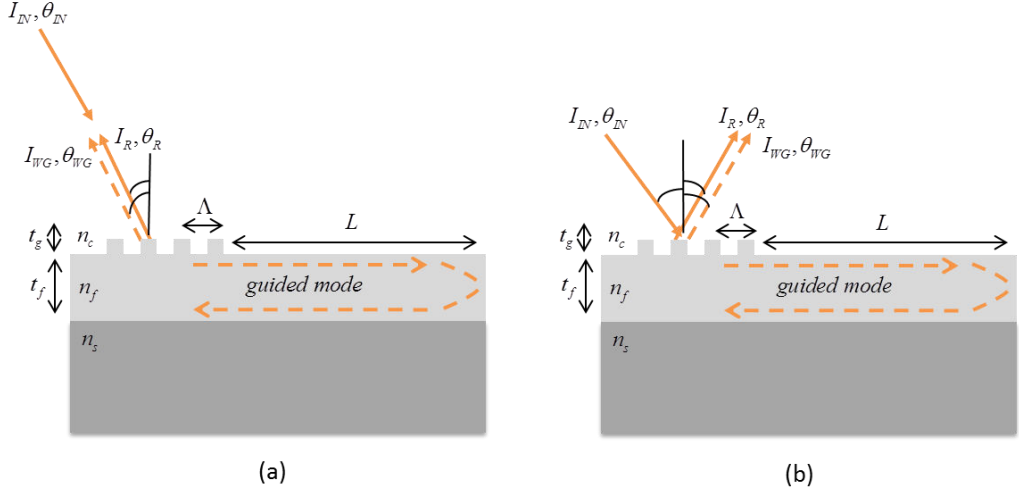

Fig. 2. Cross-section of the sensor configuration, where $t_g$ is the grating height and $\Lambda$ is the grating period. The waveguide length is $L$ and its height is $t_f$. (a) One port (Littrow-type) configuration, where *IN* represents the incoming beam, *WG* represents the beam out-coupled from the waveguide and *R* represents the beam back-diffracted directly by the grating. (b) A variation of the previous configuration, where the two out-coming beams are not directed in counter-direction of the incoming beam (non-Littrow).

Starting for reference at an ideally tuned situation, the light is affected at three stages as sensing takes place and the effective refractive index of the waveguide is slightly modified:

1. The input coupling efficiency is reduced following detuning.
2. The light propagating at the waveguide undergoes a change in phase and eventually attenuation.
3. At the output stage the resonant out-coupling angle $\theta_{WG}$ of the emerging beam *WG* is changed and its overlap with the reference beam *R* is modified.

The readout of the device is attained by recording the far-field of the emerging field composed of the coherent superposition of the beams *WG* and *R*. The formalism describing the changes at the output combination follows: Starting with the angle of the input beam $\theta_{IN}$, the reference's beam angle $\theta_R$ diffracted directly back to the cover medium is calculated from the diffraction grating equation [13]:

$$\sin(\theta_{IN}) + \sin(\theta_R) = m \frac{\lambda}{n_c \Lambda} \qquad m = 0, \pm 1, \pm 2, \ldots \qquad (1)$$

where $\Lambda$ is the grating period, m is the diffraction order provided $|\sin(\theta_R)| \leq 1$, $n_c$ is the cover refractive index and $\lambda$ is the wavelength in free space. Assuming that $n_c$ is not affected by the sensing process, the reflected beam angle $\theta_R$ does not change consequently and that beam can perform as a reference. This situation corresponds to biosensors where the effective mode index is changed by a linking process. In cases when the cover index does change, the formalism can be modified in a straightforward way.

Independently from the back-diffraction process, if we require I/O coupling of the light into and out of the waveguide we shall require that the corresponding I/O angles shall fulfil [14]:

$$\sin(\theta_{WG}^{in}) = \frac{n_{eff}}{n_c} + \frac{\lambda}{n_c \Lambda} m^{in}, \quad \sin(\theta_{WG}^{out}) = \frac{n_{eff}}{n_c} + \frac{\lambda}{n_c \Lambda} m^{out}, \qquad (2)$$

where $m^{in}$ and $m^{out}$ are the input and output coupling orders and $n_{eff}$ is the effective refractive index of the guided wave. Here, the I/O coupling angles $\theta_{WG}^{in}$ and $\theta_{WG}^{out}$ are affected by the changes in the waveguide effective refractive index during the sensing process.

The special requirement of our particular configuration is that the two processes mentioned above should take place simultaneously, meaning: $\theta_{IN} = \theta_{WG}^{in}$ and $\theta_R = \theta_{WG}^{out}$. Substituting these conditions, into Eq. (1) and Eq. (2), will translate into a single condition:

$$\frac{2\Lambda}{\lambda_{wg}} = m - (m^{in} + m^{out}) \quad , \qquad (3)$$

where $\lambda_{wg} = \lambda/n_{eff}$. This last relation is a Bragg-type condition, meaning that the grating period $\Lambda$ needs to be whole number of $\lambda_{wg}/2$. This requirement is unique to our configuration and is essential for a device that displays both effects of coupling and interference utilizing a single I/O grating. For the fully retro-reflecting structure seen in Fig. 2(a), we require in addition $\theta_{IN} = \theta_R$ (Littrow condition). Besides the propagating light's wavelength, Eq. (3) depends solely on fabrication parameters. In order to cope with possible deviations of that condition, we further analyze it and devote a tolerance analysis presented in the Appendix.

## 2.2 System far-field output analysis

The system output analysis is based on the superposition of the returning field out-coupled by the grating and the field directly diffracted by the grating. The simulation is based on following the signal field that was in-coupled by the grating according to the steps outlined in the previous section, and superposing the out-coupled field $f_{WG}(x)$ with the reference field $f_R(x)$ directly diffracted by the grating. Accordingly, assuming a Gaussian beam at the source the outbound fields at the grating's plane are expressed by:

$$f_{WG}(x) = f_{01} \exp(-\alpha_L x) \cdot \exp(j \frac{2\pi}{\lambda} n_{eff} \cdot 2L) , \quad f_R(x) = f_{02} \exp(-\frac{x^2}{w^2}) , \quad (4)$$

where $f_{01}$ and $f_{02}$ are complex constants, $\alpha_L$ is the leakage parameter of the guided mode into the free-space mediated by the grating and $((2\pi/\lambda)n_{eff} \cdot 2L)$ is the phase accumulated in the waveguide. The leakage parameter of the grating coupler was firstly calculated according to the analysis of Tamir and Peng [15], and subsequently, the width of the input Gaussian beam was determined for maximum coupling efficiency, namely [15]:

$$w = \frac{1.36}{\alpha_L \cdot \sec(\theta_{IN})} . \quad (5)$$

The expected outcome of the sensor system is simulated by projecting the far-field of the sum of the fields, $f_{WG}(x) + f_R(x)$, as expressed in Eq. (4) by means of a suitable Fourier lens and measuring the intensity spatial distribution. In the sensing process owing to changes in the waveguide effective refractive index, the guided wave phase and the nominal coupling angle ($\theta_{WG}$) changes, while the reference angle ($\theta_R$) stays constant. The difference between theses angles ($\Delta\theta$) is translated into a lateral shift $f\Delta\theta$ of the main lobes at the Fourier plane. Concurrently with the shift an interference effect takes place between these two complex fields.

## 3. Simulation results and discussion

As a model system, a waveguide based on oxidized porous silicon (PSiO$_2$) is chosen [16]. This system enables the control of the layers' refractive indices and thicknesses by means of regulating the current and time in a well-known electrochemical process [17]. In our model two PSiO$_2$ layers are formed on a silicon substrate, the upper PSiO$_2$ layer servers as the waveguide layer and the other serves as a cladding between the waveguide layer and the silicon substrate. The grating coupler is etched directly into the upper porous silicon waveguide [1,18]. In our simulation refractive indices of 1.51 and 1.28 are assumed for the core and cladding PSiO$_2$ layers respectively [19]. A wavelength of $\lambda$ = 633nm is placed for the propagating light. The waveguide is designed to supports only a single TE mode. The device parameters are indicated in Fig. 2, and hold the following values: $t_g$=200nm, $L$=1mm, $\Lambda$=0.667µm and $t_f$=395nm.

Our simulation assumes the fully back-reflecting structure seen in Fig. 2(a), meaning nominally: $\theta_{IN} = \theta_R = \theta_{WG}$, with the corresponding diffraction orders m =1, m$^{in}$ = m$^{out}$ = -1. The simulation accounts also for possible deviations from the nominal conditions due to limited resolution in the lithographic processing of the grating and inaccuracy of the rotation stage (see Appendix). Specifically a grating period deviation of 2nm and a rotation stage inaccuracy of 0.01$^0$, are assumed in the modelling. The assumed inaccuracies translate into an angular deviation between the output angles $\theta_{WG}$ and $\theta_R$. The deviation will induce a shift of the main lobes at the far field, and the two outgoing fields no longer fully overlap. In order to cancel this effect we vary the input angle by $\Delta\theta_{IN} = -0.37°$ (see Appendix Eq. (6)). This variation displaces our nominal condition from the strict Littrow situation, meaning now: $\theta_{IN} \neq \theta_R = \theta_{WG}$ at the expense of the input efficiency. At this situation the efficiency drops to 30% from its nominal value (see Appendix Eq. (7)). After validating that light is coupled into the waveguide even at non-ideal conditions, and that the directly diffracted and out-coupled beams angles overlap significantly, we simulated the sensing process by calculating and plotting the measured intensity at far-field.

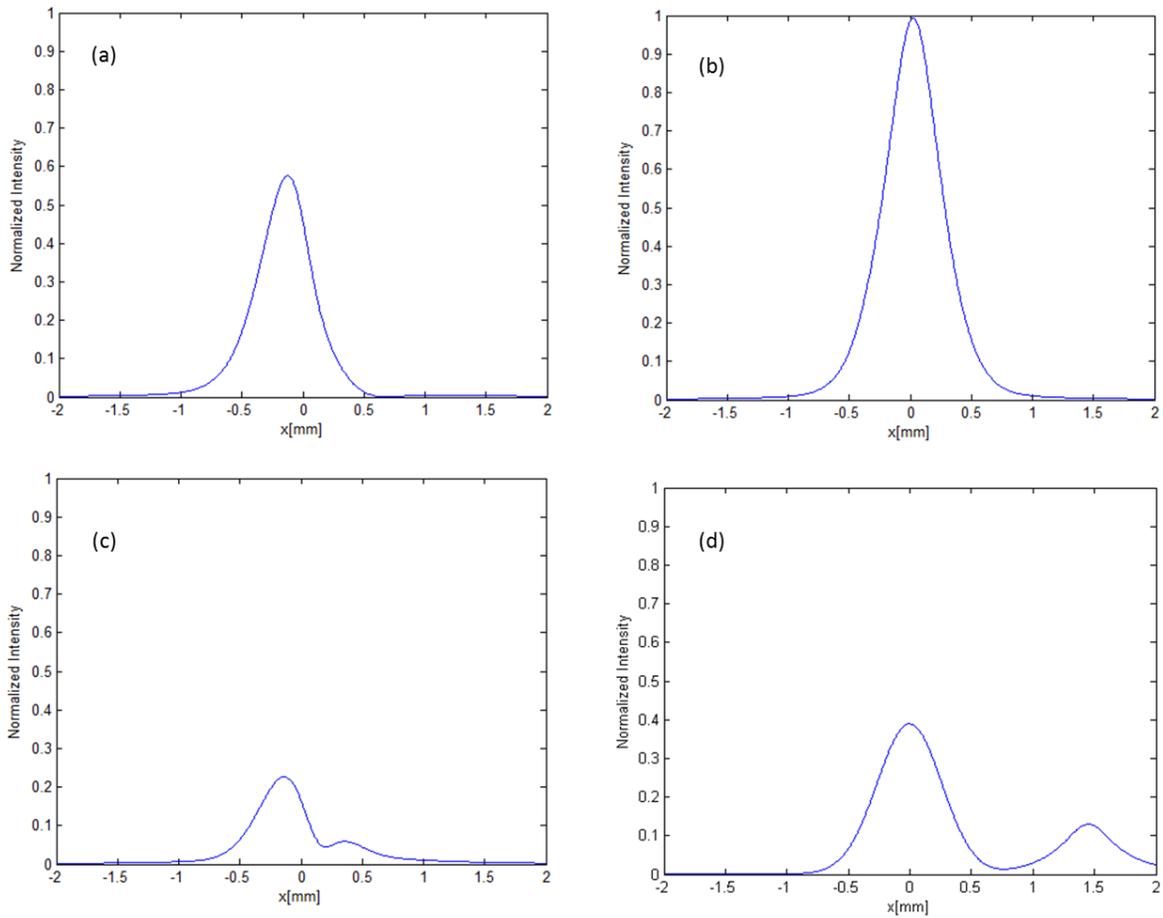

Fig. 3 Spatial intensity patterns at far field for different values of the waveguide's refractive index change: (a) $\Delta n_2 = 0$, (b) $\Delta n_2 = 3.5 \cdot 10^{-4}$, (c) $\Delta n_2 = 2.1 \cdot 10^{-3}$ and (d) $\Delta n_2 = 2 \cdot 10^{-2}$ (Media 1).

In Fig. 3(a)-(d), the far-field pictures for various changes in the waveguide refractive index are presented. The series of figures was taken from the animation in [Media 1](Media 1). The animation follows continuously the changes in spatial intensity pattern at the far-field plane as the refractive index of the waveguide varies in range of $-2 \cdot 10^{-2} \leq \Delta n \leq 2 \cdot 10^{-2}$.

As seen from Fig. 3 and from the animation, the change in shape between the patterns as the refractive index of the waveguide varies even at a slight amount is visually noticeable. Two predicted phenomena take place in conjunction: First, an angular shift between the out-coupled beam and the beam diffracted directly by the grating. Since the directly diffracted beam is the reference one, its location stays constant at the center of the frame, while the location of the second peak moves from left to right. The second effect that takes place in conjunction to the peak's displacement is a rapid amplitude change at the overlap region, as seen by comparing Fig. 3(a) and Fig. 3(b). The amplitude change is a consequence of interference between the two beams and depends on the phase difference accumulated on while propagating in the waveguide. The two effects characteristic of our sensing scheme clearly occur at widely different sensitivity scales. Another mode of monitoring the device is displayed in Fig. 4. Here the expected intensity of a single detector located at the center of the reference beam is plotted as a function of changes in the waveguide's effective refractive index. The rapid swings in amplitude reflect the interference effect while the changes in contrast of the oscillations are a consequence of the partial overlap variations between these two beams. A rough estimation of the sensitivity of the method can be extracted from this plot: For the assumed parameters, 80 oscillations swings appear for a change of $\Delta n_2 = 1.2 \cdot 10^{-2}$ meaning a sensitivity of $1.5 \cdot 10^{-4}$ RIU per swing. For an acknowledged resolution of 0.2% in power within a single oscillation for optical sensors [20] a nominal refractive index resolving power of $\Delta n_2 = 3 \cdot 10^{-7}$ RIU can be inferred.

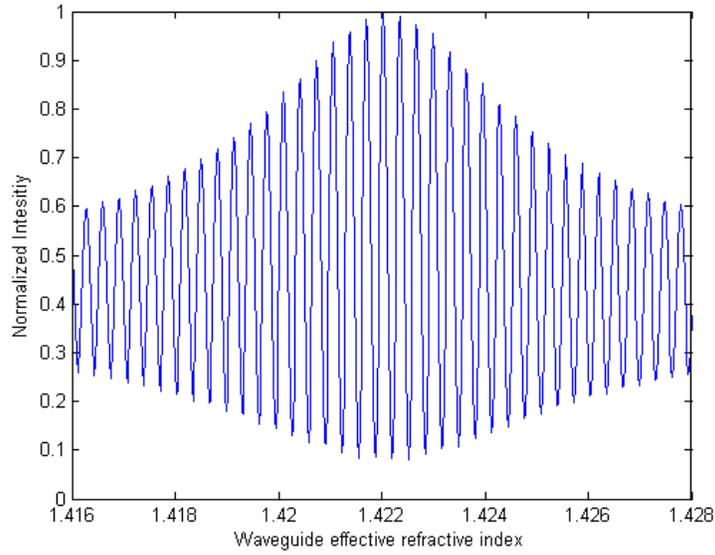

Fig. 4. Intensity vs Waveguide effective refractive index, demonstrating the sensing at two different scales. The total measurement range is about $1.2 \times 10^{-2}$ RIU while a single swing in intensity spans $1.5 \times 10^{-4}$ RIU. If a relative power resolution of 0.2% is assumed a resolving power of $3 \times 10^{-7}$ is attainable.

In order to confirm our analytical calculations, we performed FDTD numerical simulations (Lumerical Solutions) that are presented in Fig. 5. Fig. 5(a) shows the electrical power density at the waveguide's propagating section following the grating coupler. As a result of the back reflection at the end of the waveguide a standing wave is observed. As comparison, Fig. 5(b) presents the electrical power density for the standard case, where no reflection occurs at the end of the waveguide, and a uniform propagating wave is observed. In Fig. 5(c) the electric field profile at the out coupling plane is plotted together with its corresponding optimal exponential fit. The fitting furnishes a leakage parameter $\alpha = 0.02(\mu m)^{-1}$, in fair agreement with the value calculated from the analytic approximate approach of ref. [15].

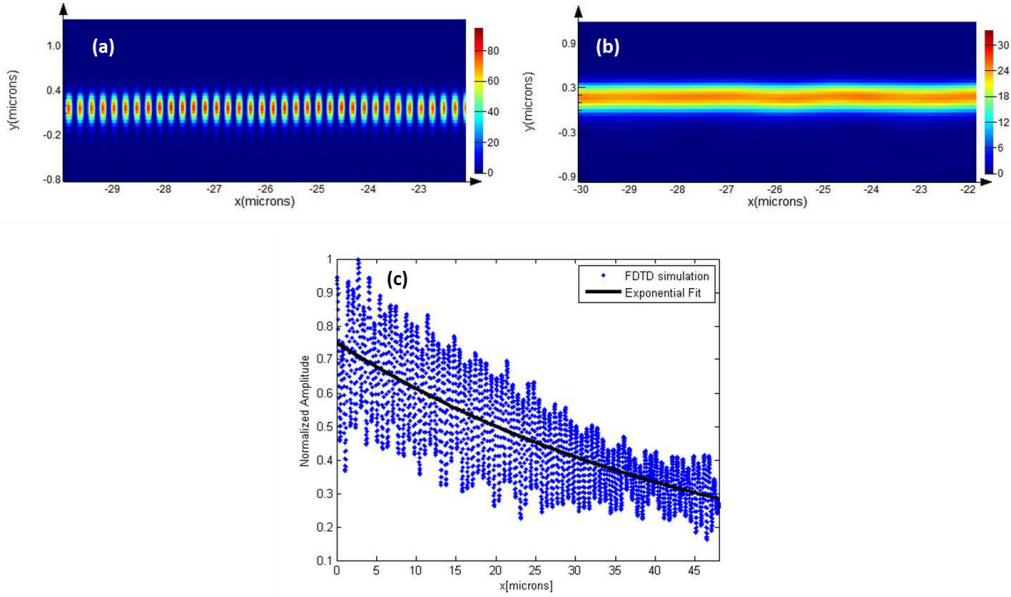

Fig. 5. FDTD simulation results. Fig. 5(a) and (b) display the electrical power density at the waveguide region. (a) In our back-reflecting configuration, and (b) In a standard configuration without reflection at the end of the waveguide. A clear standing wave is seen in the case of the back reflecting sensor. Fig. 5(c) shows the profile of the out coupled field at the grating region and its optimal exponential fit.

## 4. Conclusion

We presented and analyzed a simple waveguided back-reflecting interferometric sensor based on a single diffraction grating port. In the sensing process two effects are simultaneously measured, namely, the phase change of the propagating waveguide mode and its coupling angle change. These two effects display variations at different sensitivity scales and their simultaneous measurements provide both high sensitivity and enhanced dynamic range. The optical properties of the sensor were fully studied based on grating coupler and free-space diffraction theories. The I/O coupling processes at the grating region and the reflection at the end of the device were simulated additionally by a detailed FDTD numerical procedure. Design and fabrication tolerances considerations are presented as well in the following Appendix. Our analysis suggests that the device presented has a sensitivity compared to the best reported interferometric sensors with the additional advantages of simplicity and enhanced dynamic range. A sensor made of porous silicon was chosen as a specific example, but the results can be readily extended to any waveguided sensor were the effective refractive index changes as a result of the sensing process.

## 5. Appendix: Fabrication and alignment tolerances

The simplicity in structure and operation of the proposed device implies in the fulfillment of Eq. (3). This is a Bragg-type condition meaning that the grating period $\Lambda$ needs to be whole number of $\lambda_{wg}/2$. This requirement is not essential for conventional grating-based sensors but unique for our device, depending solely on the operation wavelength and fabrication conditions. If the optical source is wavelength-tunable the condition may be straightforwardly attained. Alternatively, since we aim for simplicity and low cost, a deviation in the fulfillment of condition (3), can be compensated by deviating the input coupling angle $\theta_{IN}$ from its nominal value. We briefly describe the derivation of the evaluation of the allowed deviation ($\Delta\theta_{IN}$) in order to compensate for fabrication inaccuracies.

In a practical alignment procedure, the grating can be rotated with respect to input beam, up to the point where the directly diffracted beam and the beam out-coupled from the waveguide overlap. Formally, referring to Eqs. (1) and (2), the required angular deviation $\Delta\theta_{IN}$ is calculated by placing $\theta_{IN} = \theta_{WG}^{in} + \Delta\theta_{IN}$, $\theta_R = \theta_{WG}^{out}$ yielding:

$$\Delta\theta_{IN} = \frac{m \cdot \dfrac{\lambda}{n_c \cdot \Lambda} - \sin\theta_{WG}^{out} - \sin\theta_{WG}^{in}}{\cos\theta_{WG}^{in}} \quad . \tag{6}$$

Once the deviation of the angle is determined, the coupling efficiency is straightforwardly calculated by the overlap integral:

$$\eta = |\int U_{WG}^{in}(x) \cdot U_{IN}(x - d_{opt}^{'}) \cdot \exp(-jk\Delta\theta_{IN} x) \cdot dx|^2 \quad , \tag{7}$$

where $U_{WG}^{in}(x)$ and $U_{IN}(x-d_{opt}^{'})$ are the normalized amplitudes of the corresponding beams at the entrance port. In Eq. (7) we further generalized the expression to include $d_{opt}^{'}$ which is the optimal displacement between both origins of both spots for maximal overlap. As pointed out in Section 3, for the assumed inaccuracies in fabrication, a rotation of $\Delta\theta_{IN} = -0.37°$ will realign the out coming beams at the expense of a reduction in coupling efficiency. The needed rotation control is thus well in the range of commercially available rotation stages.

## Acknowledgement

This research was supported in part by a grant of the Israel Science Foundation (ISF).